**NU-SETI: A PROPOSAL TO DETECT EXTRA-TERRESTRIAL SIGNALS CARRIED BY NEUTRINOS**

Ephraim Fischbach, John T. Gruenwald

Department of Physics and Astronomy, Purdue University, West Lafayette, IN 47907 USA

February 9, 2017

**Abstract**

Recent observations of changes in radioactive decay rates associated with the annual variation of the Earth-Sun distance, with solar rotation, and particularly with solar storms, suggest that radioactive decay rates may be responding to small changes in ambient neutrino/antineutrino flux. We propose to build a network of detectors (NU-SETI), based on monitoring radioactive decays, to search for pulsed signals from an extra-terrestrial source carried by neutrinos or antineutrinos.

---------------------------------------------------------

An advanced extraterrestrial civilization seeking to communicate with us might be using pulsed neutrino beams in addition to (or instead of) electromagnetic radiation. (We use "neutrino" generically to refer to any of the 6 known neutrino/antineutrino "flavors".) Since neutrinos are weakly interacting, any signal carried by a neutrino beam is less likely to be distorted en route to Earth than would be the case for an electromagnetic signal. On the other hand the very weakness of neutrino interactions makes it difficult to detect a neutrino signal using conventional techniques: Even the world's largest neutrino detector, Super-Kamiokande, detects fewer than two dozen solar neutrinos per day.

However, recent analyses of data on radioactive decays obtained by our group and others [1-6] have revealed the presence of various time-dependent deviations from the usual exponential decay law, which appear to be correlated in time with changes in the flux of solar and/or cosmic neutrinos. Specifically a major solar storm on 13 December 2006 at 2:37GMT, produced a simultaneous sharp drop in the decay rate of a $^{54}$Mn sample in our laboratory at Purdue [1]. A subsequent analysis of 2 years of data taken in our laboratory [2] has revealed a statistical correlation between the onset of solar storms and changes in $^{54}$Mn decay rates. Various characteristics of the decay data suggest that the changes in decay rates could be correlated with similar changes in neutrino flux. This conclusion was significantly reinforced in a recent paper by our group [3] in which we showed that some periodicities seen in decay rate data obtained at Brookhaven National Laboratory were very similar to those seen in neutrino data obtained by the Super-Kamiokande solar neutrino detection.

Even though we do not yet understand theoretically why some radioactive decays are particularly sensitive to neutrino-induced perturbations, the premise of our NU-SETI proposal is that this is an experimental fact, which an advanced extraterrestrial civilization might clearly understand and could be using to communicate with us. Given the aforementioned benefits of communication via neutrinos compared to radio waves, NU-SETI thus represents a new technology which could thus expand our reach in search of advanced extraterrestrial civilizations.

In searching for SETI signals carried by neutrinos, there are at least two classes of signals that might be accessible to us. We start by recognizing that we already have the capability of generating pulsed neutrino beams at Fermilab, starting from pulsed proton beams [7]. Specifically a pulsed beam was sent over a distance of 0.66 miles at an effective bit rate of 0.1 bits/sec, and was received with a detection accuracy of 99%. If we assume an advanced civilization can do somewhat better, then we can search for "universal" strings of pulses, say, those characterizing prime numbers 1,2,3,5,7,… The other class of signals would be those specific neutrino signals associated with an advanced civilization running exclusively on fission or fusion sources [8] all of which produce characteristic neutrino signals.

In practical terms our proposed NU-SETI system would be a scalable array of individual sites spread over the world, each measuring the decay rate(s) of one or more specifically chosen radioactive sources, such as $^{54}$Mn. We estimate that each site could be set up for approximately $20K, so that $20M would fund approximately 1000 sites. These sites would be connected to a common central server which would correlate the incoming data using extensions of software we have already developed. This software will be enhanced with "machine learning capabilities" to further enhance the sensitivity of the distributed monitoring system. NU-SETI will be particularly attractive from a cost point of view, since radioactive decay data collected by this array could also be used to predict solar storms (and thus mitigate their effects) using technology which our group has already patented [9]. Predicting solar storms, using data collected by NU-SETI, would thus likely motivate investment in NU-SETI by sectors sensitive to the effects of solar storms such as electric power companies and the military.